\documentclass[aps,pra,superscriptaddress,a4paper,reprint]{revtex4-1}

\usepackage[utf8]{inputenc}
\usepackage{amsmath}
\usepackage{graphicx}
\usepackage[left=2cm,right=2cm,
    top=2cm,bottom=2cm,bindingoffset=0cm]{geometry}
\usepackage{ulem}

\begin{document}

\title{Optical trapping of nanoparticles by full solid-angle focusing}

\author{Vsevolod Salakhutdinov}
\affiliation{Max Planck Institute for the Science of Light,
  Guenther-Scharowsky-Str. 1/ building 24, 91058 Erlangen, Germany}
\affiliation{Friedrich-Alexander-Universit\"at Erlangen-N\"urnberg (FAU),
  Department of Physics, Staudtstr. 7/B2, 91058 Erlangen, Germany}

\author{Markus Sondermann}
\email[]{markus.sondermann@fau.de}
\affiliation{Max Planck Institute for the Science of Light,
  Guenther-Scharowsky-Str. 1/ building 24, 91058 Erlangen, Germany}
\affiliation{Friedrich-Alexander-Universit\"at Erlangen-N\"urnberg (FAU),
  Department of Physics, Staudtstr. 7/B2, 91058 Erlangen, Germany}

\author{Luigi Carbone}
\affiliation{CNR NANOTEC-Istituto di Nanotecnologia U.O. Lecce,
  c/o Polo di Nanotecnologia-Campus Ecotekne, via Monteroni, 73100 Lecce, Italy}

\author{Elisabeth Giacobino}
\affiliation{Laboratoire Kastler Brossel, UPMC-Sorbonne Universités,
  CNRS, ENS-PSL, Research University, Collège de France, 4 place
  Jussieu,case 74 F-75005 Paris, France}
\affiliation{Max Planck Institute for the Science of Light,
  Guenther-Scharowsky-Str. 1/ building 24, 91058 Erlangen, Germany}

\author{Alberto Bramati}
\affiliation{Laboratoire Kastler Brossel, UPMC-Sorbonne Universités,
  CNRS, ENS-PSL, Research University, Collège de France, 4 place
  Jussieu,case 74 F-75005 Paris, France}

\author{Gerd Leuchs}
\affiliation{Max Planck Institute for the Science of Light,
  Guenther-Scharowsky-Str. 1/ building 24, 91058 Erlangen, Germany}
\affiliation{Friedrich-Alexander-Universit\"at Erlangen-N\"urnberg (FAU),
  Department of Physics, Staudtstr. 7/B2, 91058 Erlangen, Germany}
\affiliation{Department of Physics, University of Ottawa, Ottawa,
  Ont. K1N 6N5, Canada}

\date{\today}

\begin{abstract}
Optical dipole-traps are used in various scientific fields,
including classical optics, quantum optics and biophysics.
Here, we propose and implement a dipole-trap for nanoparticles that is based
on focusing from the full solid angle with a deep parabolic mirror.
The key aspect is the generation of a linear-dipole mode which is
predicted to provide a tight trapping potential.
We demonstrate the trapping of rod-shaped nanoparticles and validate
the trapping frequencies to be on the order of the expected ones.
The described realization of an optical trap is applicable for various
other kinds of solid-state targets.
The obtained results demonstrate the feasibility of
optical dipole-traps which simultaneously provide high trap
stiffness and allow for efficient interaction of light and
matter in free space.
\end{abstract}

\maketitle

\section{Introduction}

About fifty years ago optical forces have been used for the first
time in trapping and localizing particles\,\cite{ashkin1970}. Since
then, optical traps have become a workhorse in many fields of
science, not least in atomic physics, quantum optics and
optomechanics. In these areas, experiments are typically conducted
in the Rayleigh regime, where the size of the trapped particle is
much smaller than the wavelength of the trapping beam. In this case,
the particle acts as a point-dipole, and the trapping potential
scales with the product of the particle's polarizability and the
squared modulus of the spatially varying amplitude of the electric
field.

Choosing a specific particle, the depth of the trapping potential is
maximized by maximizing the field amplitude of the focused trapping
beam. The optical modes yielding the maximum possible field
amplitude at constant input power are electric-dipole
modes\,\cite{basset1986}. These modes  have been suggested
previously to maximize the interaction of light and single atoms in
free space~\cite{quabis2000,vanenk2004,lindlein2007,sondermann2007}. 
In this paper, we demonstrate for the first time a dipole trap
generated with such a mode.

The key aspect in the generation of an electric-dipole wave is the
focusing of a suitably shaped light mode with optics covering the
entire solid angle.
Such an optical element can be realized with a parabolic mirror (PM)
that is much deeper than its focal length\,\cite{lindlein2007}.
As outlined in the next section, optical tweezers based on a dipole
wave not only result in a deeper trapping potential, but also in high
trap frequencies.

There are several current research topics which could benefit from
an optical trap realized with a deep PM. Recently, there has been
experimental progress in the trapping of nanoparticles in the
context of cavity-free
opto-mechanics\,\cite{li2011,neukirch2013,arita2013,gieseler2013,millen2014,neukirch2015},
where stiffer traps could help in reaching the quantum regime of the
particle's motion\,\cite{rodenburg2016}. Another motivation is the
application of the free-space light-matter coupling scheme proposed
in Ref.\,\cite{sondermann2007} for solid-state quantum targets.
By trapping such targets in an optical dipole-trap, the actual
excitation of the freely levitated quantum emitter would not suffer
from disturbances induced by refractive index boundaries as present
when focusing onto emitters located in a solid-state host medium.
For example, Ref.~\,\cite{vamivakas2011} discusses the detrimental
effects induced by boundaries between different dielectrics when
focusing onto buried InAs semiconductor quantum dots. Furthermore,
the fluorescence collected by a deep PM from a single photon source
can be extremely high, as exemplified for a single ion effectively
radiating as an isotropic point source\,\cite{maiwald2012}. For a
linear-dipole emitter oriented along the PM's axis, the collection
efficiency can be close to unity even for a PM of finite depth.

In the next section, we describe the basic features of optical
tweezers implemented by focusing via a deep PM, highlighting its
favorable figures of merit.
Section\,\ref{sec:Exp} reports on the experimental implementation
of a PM dipole trap.
Finally, we provide an outlook to future work.

\section{Concept of a deep parabolic mirror dipole trap}
\label{sec:Concept}

\begin{figure}[tb]
  \centering
      \includegraphics[width=8cm]{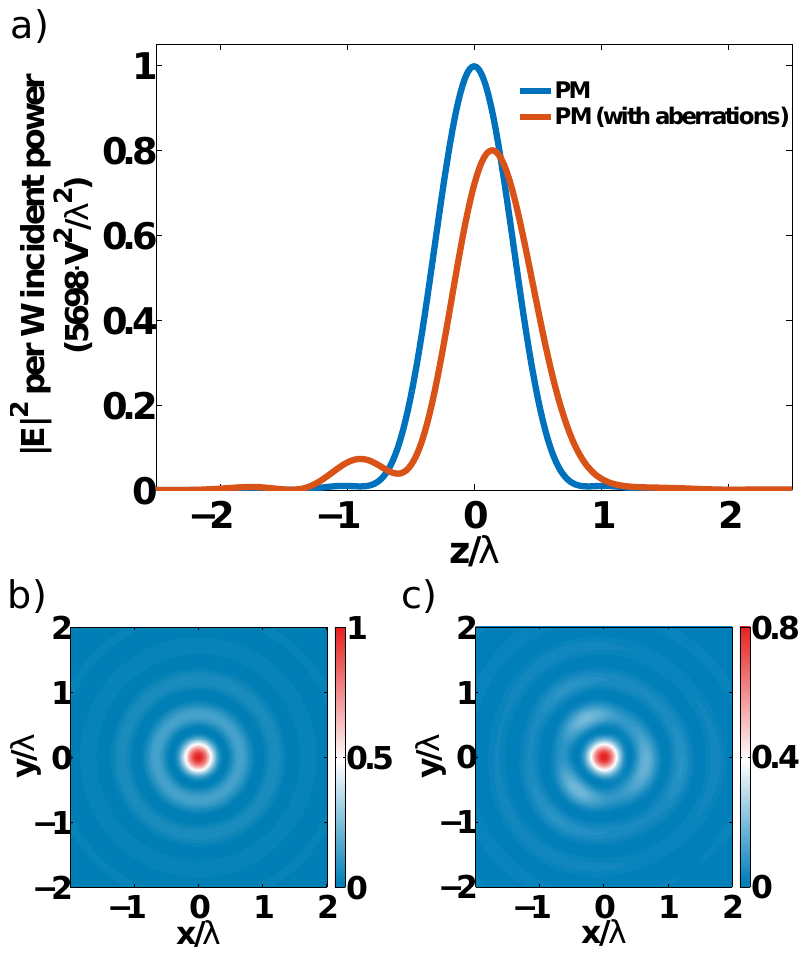}
      \caption{\label{fig:Simulation}
    Simulated intensity distributions for a dipole trap based on a deep PM
    (see text for all parameters).
    The PM is illuminated by a radially polarized doughnut mode.
    (a) Intensity distribution along the optical axis of the PM.
    The blue curve denotes the case of an aberration free
    mirror, the orange curve is based on a simulation accounting for the
    aberrations of the mirror used in the experiments and a wavelength
    $\lambda=1064\,\textrm{nm}$.
    (b,c) Intensity distribution in the plane perpendicular to the
    optical axis at the axial position with highest intensity for the
    diffraction limited case (b) and accounting for aberrations (c).
 }
\end{figure}

In the past few years spherical mirrors and parabolic mirrors (PMs)
have been considered in the literature and used for trapping micro-
and nano-objects. In Ref.\,\cite{merenda2007} an array of parabolic
mirrors with half opening angles of 38$^\circ$ has been used to trap
particles in a fluid. Analogously micro-fabricated spherical mirrors
have been proposed for the implementation of dipole traps for
neutral atoms\,\cite{goldwin2008}, whereas the experimental
implementation of a single such trap was demonstrated
recently\,\cite{roy2012}. However, also there the solid angle
spanned by the trapping mirror was still fairly small.

In this article we analyze and use a PM design with a depth much
larger than the focal length. The mirror geometry is the same as
used in Refs.\,\cite{maiwald2012,fischer2014}. A focal length of
$f=2.1\,\textrm{mm}$ and an aperture radius of 10\,mm result in a
half-opening angle of $135^\circ$ and a depth of the parabola of
11.9\,mm.
This yields a coverage of 94\% of
the solid angle relevant for a linear dipole oriented along the axis
of the PM. To produce a dipole wave, the PM is illuminated with a
radially polarized mode with the amplitude distribution $A(r)\sim
r\cdot \exp(-r^2/w^2)$ (so-called \lq doughnut\rq\ mode). When
choosing the beam radius to be $w=2.26f$ one obtains a close-to
ideal dipole wave after reflection of the doughnut mode at the PM's
surface\,\cite{sondermann2008}. The advantages of radially polarized
modes for optical trapping -- a tighter potential in radial
direction and the absence of scattering forces on the optical axis
-- have been discussed earlier\,\cite{zhan2004}, although
not in the context of a dipole mode.

We now describe the benefits of an optical dipole trap based upon
focusing with a deep PM. The trapping potential is given by the
relation $U(\vec{q})=-\alpha/2\cdot
|E(\vec{q})|^2$\,\cite{novotny-hecht2012}, where $\alpha$ is the
polarizability of the particle and $E(\vec{q})$ the spatially
varying amplitude of the electric field used for trapping at
position $\vec{q}=(x,y,z)$. As outlined above, by increasing the
maximum field amplitude $E(\vec{q})$, the depth of the trapping
potential increases, thus achieving the maximum potential depth in
free space with an electric-dipole wave\,\cite{basset1986}.

As discussed in Ref.\,\cite{sondermann2008}, the energy fraction of
dipole radiation contained in a focused optical beam is limited by
the solid angle covered by the focusing optics. More precisely, the
relevant quantity is the solid angle
$\Omega=\int\,D(\vartheta)\sin\vartheta\,d\varphi d\vartheta$
obtained by weighting the solid angle used for focusing with the
desired angular dipole radiation pattern
$D(\vartheta)=\sin^2\vartheta$ for a linear dipole. By focusing from
full solid angle $\Omega$ is maximized to
$8\pi/3$\,\cite{sondermann2008}. For the PM used here
$\Omega=0.94\cdot8\pi/3$. According to Ref.\,\cite{sondermann2008}
one has $|E(0)|^2\propto \Omega\eta^2$ with $0\le\eta\le1$
parametrizing the similarity of the incident radiation pattern with
a dipole wave and the mirror's focus located at $\vec{q}=0$. For the
doughnut mode used here one has $\eta=0.98$. In other words, for a
given power the PM design used here can deliver 90\% of the maximum
possible potential depth (cf. Ref.\,\cite{golla2012}).

Next we estimate the achievable trapping frequencies.
We compute the focal intensity distribution for our chosen mirror
geometry and incident beam, assuming an aberration free PM.
The simulations were performed using a generalization of the
method by Richards and Wolf\,\cite{richards1959}.
The results are reported in Fig.\,\ref{fig:Simulation}(a) and (b).
In the vicinity of its minimum, the trapping potential can be
approximated by the harmonic potential $U(\vec{q})\approx k\cdot
q^2/2$, where $k= m\omega^2$ is the trap stiffness, $m$ is the trapped 
particle's mass and $\omega$ is the trap frequency.
$k$ is obtained from fitting axial and
radial cuts through the focus to the harmonic potential $U(z)$ and
$U(r)$, respectively.
We obtain $k_z = 5/\lambda^2\cdot\alpha|E(0)|^2$ for the trap
stiffness in axial direction and
$k_r=15/\lambda^2\cdot\alpha|E(0)|^2$ in radial direction.
A possible influence to the trapping potential of the
polarization components perpendicular to the optical axis can be
neglected due to their low magnitude in relation to the longitudinal
field.
The transverse field components reach a maximum value of about
$3\%$ of the focal intensity approximately at 0.6\,$\lambda$
distance to the focus. Along the optical axis the off-axis field
components vanish due to symmetry reasons when focusing a radially
polarized beam.

To gauge our results, we compare the stiffnesses obtained for the deep
PM to the ones obtained  
when focusing a linearly polarized fundamental Gaussian mode with a
lens to a beam waist $w$ and a Rayleigh length $z_\textrm{R}$.
For such a mode the trap stiffnesses
are derived in Ref.\,\cite{gieseler2013} to be
$k_{z,\textrm{Gauss}}=\alpha|E(0)|^2/z_\textrm{R}^2$ and
$k_{r,\textrm{Gauss}}=\alpha|E(0)|^2/w^2$.
In an attempt to compare to a typical example we choose
$\lambda=1064\,\textrm{nm}$ and the beam parameters from
Ref.\,\cite{gieseler2013} which are $w\approx 0.54\,\mu\textrm{m}$ and
$z_\textrm{R}\approx 1.36\,\mu\textrm{m}$, obtained with an NA=0.8
objective.
The resulting trap stiffnesses are lower by a factor
of 4 and 8 in radial and axial direction, respectively,
as compared to what we obtain for the deep PM.
Moreover, this calculation assumes the same maximum focal intensity
$|E(0)|^2$ for the focused Gaussian beam and for the PM.
When comparing the stiffnesses for the same incident power in the two
cases, the improvement for the deep PM is even more pronounced,
since the fraction of the dipole weighted solid angle covered by an
NA=0.8 objective is only about $\Omega=0.25\cdot8\pi/3$ and the
maximum focal intensity is increased by a factor of 3.6 
when using the deep PM.
In total, this suggests that with a deep PM the trap stiffness
can be improved by up to one order of magnitude in the ideal case.
Other measures to improve the stiffness such as using a standing wave
trap or a cavity are not considered here as we do not want to change the
spectrum of modes of free space, retaining the highest possible
frequency bandwidth for coupling to a trapped quantum target. 

Experimentally, CdSe/CdS dots-in-rods (DRs) \,\cite{pisanello2010}
were trapped in air at normal ambient pressure. For estimating the
polarizability of these particles we neglect the spherical CdSe
core, which makes a negligible contribution to the total volume. The
DRs used in the experiment have a cylindrical shape with a length of
58\,nm and a diameter of 7\,nm.
Within good approximation, these particles can be treated as
point-like, since in all spatial directions their extent is much
smaller than the wavelength of the trapping laser.
Furthermore, using the same wavelength even larger particles have been
treated successfully as point-like scatterers in recent
literature\,\cite{neukirch2013,gieseler2013,rodenburg2016}.
Since there is no closed analytic
expression for the polarizability of a dielectric cylinder, we
approximate the CdS rods as prolate spheroids. An expression for the
polarizability of prolate spheroids is found in e.g.
Ref.\,\cite{mihaljevic2014}. For a refractive index of CdS at
1064\,nm of $n_\textrm{CdS}=2.344$ we arrive at a polarizability
$\alpha=5.28\cdot10^{-35}\,\textrm{Cm}^2/\textrm{V}$. From this
value we expect a potential depth of $|U(0)|\approx
\textrm{k}_\textrm{B}\cdot9600\,\textrm{K}$ for an incident power of
1\,W. Under ambient conditions, the trapped particle thermalizes to
room temperature due to collisions with air
molecules\,\cite{gieseler2013}. Hence, we expect that the minimal
power for trapping is about 32\,mW.

For completeness  we note that also the excitonic optical transition
has a finite polarizability. Assuming a center wavelength of
$\lambda_\textrm{exc}=595$\,nm and a radiative life time of
$\tau=15$\,ns as found for the DRs used here, we calculate an
excitonic polarizability of
$\alpha_\textrm{exc}=1.7\cdot10^{-39}\,\textrm{Cm}^2/\textrm{V}$
when focusing at 1064\,nm \cite{transNote}.
This polarizability is four orders of
magnitude below the one estimated for the rod material and can
therefore be neglected.

\section{Experimental realization}
\label{sec:Exp}

\begin{figure}[tb]
  \centering
  \includegraphics[width=8cm]{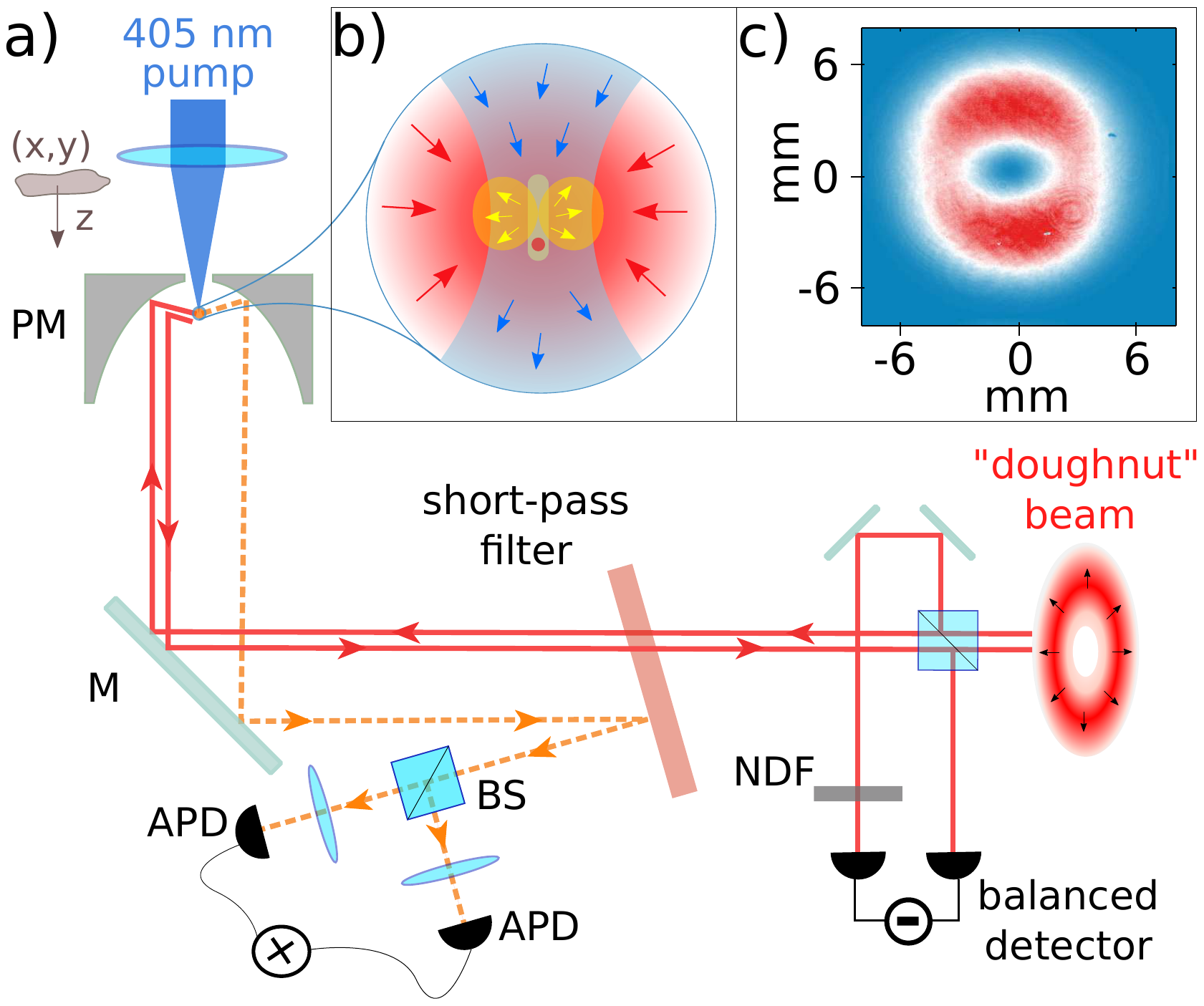}
  \caption{\label{fig:setup}
    (a) Scheme of the experimental setup.
    A radially polarized \lq doughnut\rq\ mode
    ($\lambda=1064\,\textrm{nm}$, intensity profile in panel (c)) is
    focused by a deep PM.
    Fine tuning of the beam's angular alignment is achieved by a
    mirror (M).
    The trapped nanoparticles are excited by a pulsed 405\,nm laser
    (1\,MHz repetition rate, average power $\sim 4.5\,\mu\textrm{W}$), which
    is focused onto the trapped particles through a bore hole at the PM's
    vertex, see panel (b) for illustration.
    The photons emitted by the dot-in-rod nanoparticles (595\,nm
    center wavelength) are collimated by the PM and split from the
    infrared light by a  reflective short pass filter (776\,nm
    cutoff wavelength).
    A beam splitter (BS) directs the fluorescence photons onto two
    avalanche photo diodes (APD) for detecting the presence of trapped
    particles and measurements of the second order
    intensity-correlation function. Both APDs are equipped with
    chromatic filters blocking the 405\,nm excitation light.
    The trapped particles' motion is analyzed by means of a
    balanced detector, which delivers the difference signal between the
    infrared light leaving the parabolic mirror and the incident
    light. The detection is calibrated by means of a neutral-density
    filter (NDF) such that the difference signal is canceled out when
    no particles are trapped.
}
\end{figure}

In the experiment we use an aluminum PM made by diamond turning
(Fraunhofer IOF, Jena), fabricated with the  dimensions given above.
 The surface of the mirror exhibits deviations from a
parabola. These surface deviations are measured by
interferometry\,\cite{leuchs2008}. In general, such deviations
distort the focus, leading to a reduced depth of the optical
trapping potential as well as reduced curvatures of the potential.
In order to estimate the influence of these deviations, we performed
simulations of the focal intensity distribution based on the
measured aberrations.
The corresponding phase front errors have been incorporated in
the simulations by attributing them to the phase front of the
incident beam.
The results for a wavelength of 1064\,nm are
given in Fig.\,\ref{fig:Simulation}(a) and (c). The maximum
intensity in the focal region and consequently the potential depth
is reduced by 20\%. 
Accordingly, the minimum power required to trap a single nanorod at
room temperature increases to 39\,mW.
The intensity maximum is shifted along the
optical axis of the PM by a negligible amount.
The curvatures of the trapping potential are slightly reduced to
values of $k_r=13.7/\lambda^2\cdot\alpha E_{\textrm{max}}^2$ and
$k_z=3.3/\lambda^2\cdot\alpha E_{\textrm{max}}^2$, with
$E_{\textrm{max}}$ the field amplitude at the intensity maximum in the
presence of aberrations.
The rotational symmetry
of the intensity distribution in the plane perpendicular to the PM's
optical axis, and hence the symmetry of $U(r)$, is hardly disturbed.

The experimental setup is outlined in Fig.\,\ref{fig:setup}a. A
single-mode continuous wave laser operating at 1064\,nm (Cobolt
Rumba, 2\,W output power) is sent through a liquid-crystal based
polarization converter (ArcOptix) and a spatial filter in order to
produce a radially polarized doughnut mode.
The spatial filter comprises a telescope with a pinhole
positioned in the focal plane. The pinhole rejects unwanted
higher-order modes and cleans the phase front of the transmitted
mode to a large extent.
The intensity distribution of the resulting doughnut mode is shown in
Fig.\,\ref{fig:setup}(c).
The rotational symmetry of the mode is slightly perturbed.
In order to assess the influence of this perturbation,
we characterized the generated doughnut mode by a spatially resolved
measurement of the Stokes parameters. Using the same methods and
analysis as 
described in Ref.\,\cite{golla2012} an overlap $\eta=0.95$ with an
ideal dipole-mode is obtained.
This value is in good agreement with the value assumed in our
calculations above and demonstrates a high quality of the radially
polarized mode.
The doughnut mode incident onto the PM has
a beam radius $w\sim4.7\,\textrm{mm}$.

The DR nanoparticles are delivered to the trapping region with an
approach adapted from
Refs.\,\cite{neukirch2013,millen2014,minowa2015}. The DR particles
are dispersed in toluene and diluted to a concentration of about
$10^{-9}\,\textrm{mol/l}$. In a next step this solution is dissolved
in methanol with a ratio of 1:100. The resulting solution is
sonicated for ten minutes and then transferred into an ultrasonic
nebulizer (Omron), from where it is sprayed to the vicinity of the
rear side of the PM. Droplets containing DRs diffuse in the air
towards the trapping region through an opening at the vertex of the
parabola. DRs are then trapped in the focus of the PM.

\begin{figure}[tb]
  \centering
  \includegraphics[width=8cm]{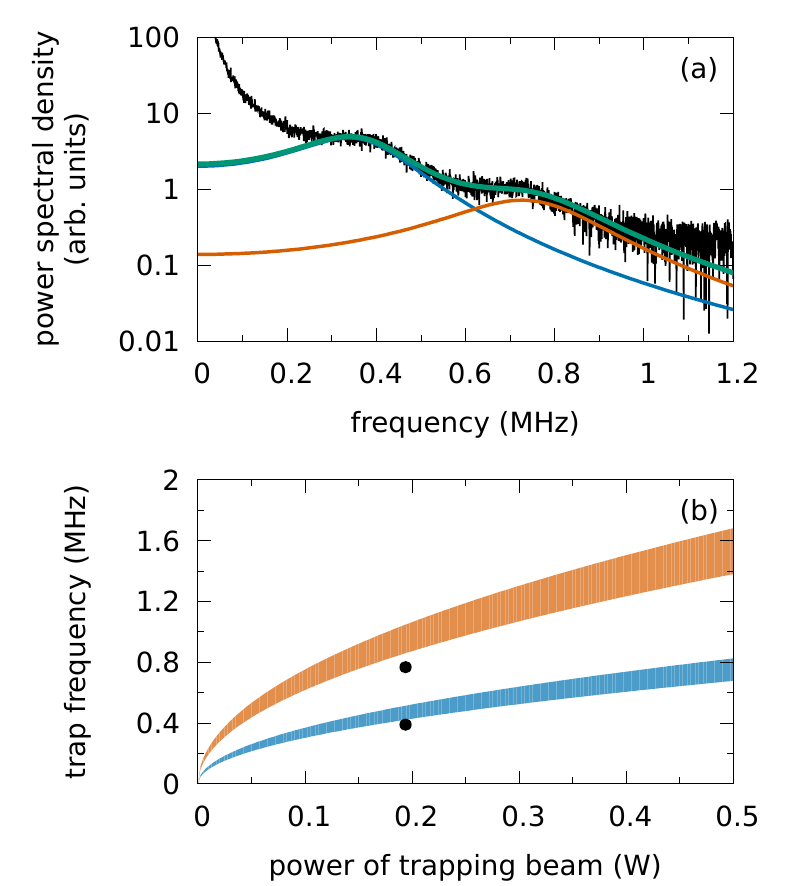}
  \caption{\label{fig:motion}
    (a) Power spectral density of the
    dynamics of the motion of a trapped cluster of DR particles
    ($g^{(2)}(0)=0.9$) 
    obtained for a trapping beam power of 194\,mW.
    Black line: experimental data after subtracting the spectral
    density measured with an empty trap.
    Green line: Double-Lorentzian fit. Frequencies below 0.3\,MHz have
    been excluded from the fitting procedure.
    Blue and orange line: Lorentzians attributed to the motion
    along the PM's optical axis and perpendicular to it, yielding
    $\omega_z/2\pi=0.389$\,MHz and $\omega_r/2\pi=0.767$\,MHz,
    respectively.
    (b) Trap frequencies as a function of the trapping beam's power.
    Shaded areas: Theoretical expectations for
    $\omega_r/2\pi$ (orange) and $\omega_z/2\pi$ (blue).
    The upper boundary of each
    area is obtained using the polarizability of a prolate
    spheroid. The lower border is obtained using the polarizability of
    a sphere.
    Both curves account for the aberrations of the PM used in the
    experiments as well as the experimentally obtained overlap of the
    doughnut mode with an ideal dipole mode.
   Black dots: Trap frequencies obtained from the spectrum in (a).
  }
\end{figure}

To detect the motion of the particles in the trap we adapt an
approach from Ref.\,\cite{mestres2015} based on the interference of
the light of the trapping laser and the light scattered from the
trapped object.
The PM reflects a huge portion of the incident light back into the
direction of incidence.
Simultaneously the PM collimates almost the entire light scattered
by the trapped particle.
Due to the motion of the particle in the trap, the resulting
interference signal is varying in time.
Since a deep PM acts as an efficient retro-reflector, the back
reflected light is much stronger than the scattered counterpart,
resulting in a small modulation due to interference on top of a
large constant background.
Hence, we apply balanced detection by subtracting a reference
signal picked off from the trapping laser before it enters the PM,
cf. Fig.\,\ref{fig:setup}.
This allows to measure the trap frequencies as shown below (see
also Fig.\,\ref{fig:motion}a).

The DR particles are optically excited by a pulsed laser operating
at 405\,nm wavelength (Alphalas, Picopower). The pulse duration of
~80\,ps is short in comparison to the excitonic life time of about
15\,ns, see also Ref.\,\cite{pisanello2010}. After initial trapping
with a power of 280\,mW, the power of the trapping laser is
generally decreased in order to reduce photo bleaching of the DRs.
It was found that the DR particles are also excited by the trapping
laser via two photon absorption. The final trap beam power of
~8.5\,mW was chosen such that fluorescence induced by two-photon
absorption alone is indistinguishable from the noise floor. 
Under these conditions DRs remained in the trap for a few tens of
minutes.
Note that the value of 8.5\,mW is smaller than the 39\,mW value at
which we expect to loose the DRs from the trap according to the
estimation above. 
This aspect is discussed at the end of this section.

At room temperature the fluorescence spectrum of the DR particles is
centered at 595\,nm. Fluorescence photons are collimated by the PM
(72.4\% reflectivity) and separated from the
retro-reflected trapping beam and from the pulsed laser by a
dichroic mirror. They are detected by two avalanche photo-diodes
(LaserComponents, Count 100) working in the Geiger mode. Bandpass
filters in front of these diodes (not shown in
Fig.\,\ref{fig:setup}) block residual light stemming from the trapping
and excitation lasers. 
Overall the net transmission via the PM to the
avalanche diodes is 26\%, including the losses at the PM's surface
and at all other optical components as well as the avalanche diodes'
quantum efficiency. Photon detection events are recorded by a
time-to-digital converter (quTau from quTools, Munich). From these
data we compute the second-order intensity correlation function
$g^{(2)}(\tau)$, by which we verify whether a single DR particle is
emitting.

\begin{figure}[tb]
  \centering
      \includegraphics[width=8cm]{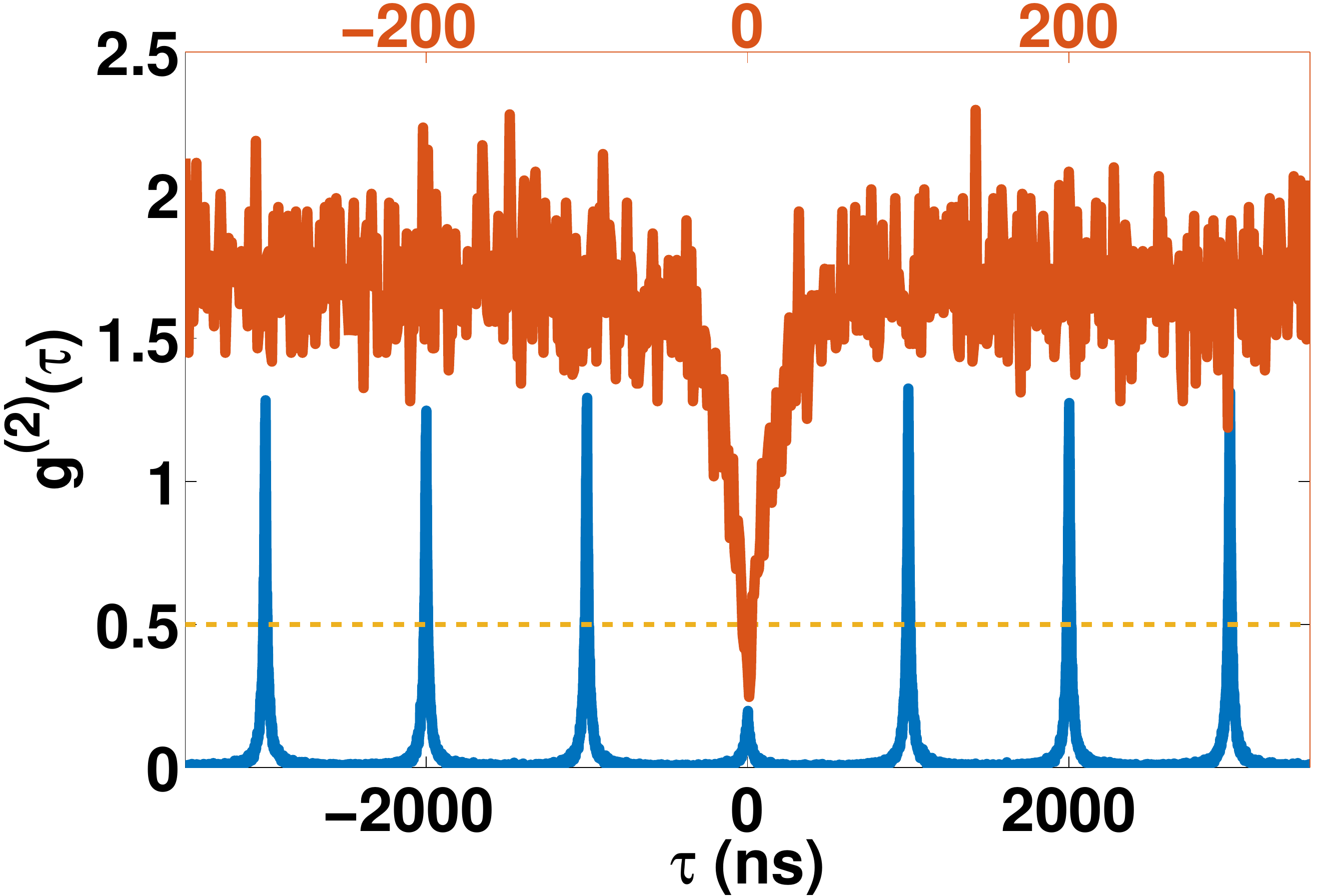}
  \caption{\label{fig:g2}
    Second order correlation functions for a single 
    CdSe/CdS nanoparticle. The orange curve corresponds to the
    function measured when exciting the nanoparticle only with the
    infrared CW trapping beam at a power level of 370\,mW.
    The blue curve is the correlation function obtained when exciting
    the nanoparticles with the pulsed 405\,nm laser and a trapping
    beam power of 8.5\,mW.
    The dashed line marks $g^{(2)}(\tau)=0.5$.}
\end{figure}

Figure\,\ref{fig:g2} shows an example of a
$g^{(2)}$ function recorded from the fluorescence of a single DR.
For excitation with the trapping beam laser only, as well as with
the pulsed laser, the normalized $g^{(2)}$ function exhibits clear
anti-bunching. When pulsed excitation is applied, the trap laser is
attenuated to power levels such that no fluorescence due to
two-photon absorption is observed. Noteworthy, in both kinds of
measurements one obtains $g^{(2)}(0)\approx0.2$. 
This imperfect antibunching for a single DR is attributed to a small
but finite probability for the recombination of more than one
exciton\,\cite{vezzoli2013}. For delays outside the central
anti-bunching dip, the $g^{(2)}$ functions are larger than one. For
the DRs this effect is attributed to blinking stemming from the
presence of a charged state\,\cite{manceau2014}. At large delays,
the measured $g^{(2)}(\tau)$ approaches unity (not shown in
Fig.\,\ref{fig:g2}).

In the experiment $g^{(2)}$ functions with values of
$g^{(2)}(0)>0.5$ are obtained very frequently, giving the
indication that clusters made of DRs are trapped. 
This can be explained by the fact that methanol, which was used to
dilute the DR solution favors DRs aggregation and cluster formation
\cite{methanolNote}.
In view of this it is most likely that the
single photon emitters studied above are also made of small clusters
of DRs, with only one DR emitting single photons, the other ones
being damaged and not emitting anymore.

This clustering effect can also explain the difference between the
calculated minimal trapping power and the significantly lower power at
which particles could be trapped stably, because a
cluster has a larger overall polarizability than a single DR. 
If one assumes that clusters are made of at least 5 to 7 DRs the low
observed trapping power is in good agreement with the calculated one
\cite{powerNote}.

Finally, we discuss the trap frequencies achieved in the experiments
with the PM based optical trap. 
The frequencies are identified by fitting a sum of two Lorentzians
$S(f)=\sum_{i=r,z}A_i/[(f^2-f_i^2)^2 + f^2\gamma_i^2]$
to power spectra of the signal acquired
with the balanced detector, cf. Fig.\,\ref{fig:setup}. 
Here, $f_i=\omega_i/2\pi$ represent the trap frequencies, the
$\gamma_i$ denote damping terms and the $A_i$ are used as fit
constants (see e.g. Ref.\,\cite{li2011}). 
Figure\,\ref{fig:motion}(a) shows the outcome of such a procedure for the
case of trapping a cluster.
The resulting trap frequencies are compared to the ones expected from
calculations in Fig.\,\ref{fig:motion}(b).
The experimental values are about 10\% lower than the ones calculated
when assuming a spherical shape.
A sphere is likely to be a good approximation for the geometry of a
cluster of particles.
Nevertheless, one cannot ensure perfect spherical symmetry nor
guarantee a homogeneous distribution of mass within a cluster.
Therefore, witnessing some deviations in the experiment
is not surprising.
Another potential origin of discrepancies on the few-percent level
could be found in the residual aberrations of the optical elements in
the beam path. 
The deviation of 25\% to the trap frequency for a prolate spheroid is more
pronounced.
On the other hand, the ratio $f_r/f_z\approx2$ is the one expected from
$\sqrt{k_r/k_z}$ when taking into account the curvatures of the focal
intensity distribution in the presence of the aberrations of the PM.
We therefore conclude that our optical trap has a high stiffness,
with performances in good agreement with our model.

\section{Outlook}
\label{sec:Discuss}

We have demonstrated the trapping of nanoparticles in a deep PM via
an optical dipole trap, together with  the collection of
the light emitted by the trapped particles.
This result shows the feasibility of an efficient interaction between
solid state targets and the light field in free space,
and opens the way to the study of solid-state quantum
objects without any disturbance induced by a host medium.
As a next step, one can envision a setup incorporating two dipole
waves: one for trapping the particle, the
second one for driving the optical transition of interest.

As test particles we have used CdSe/CdS dot-in-rods.
These objects are known to emit with a radiation pattern strongly
resembling a linear dipole\,\cite{pisanello2010}.
Therefore, we expect high efficiencies in collecting the fluorescence
photons emitted by these DRs with a deep PM.
Besides that, the linear-dipole property of the excitonic transitions
of the used DRs makes them suitable candidates for efficient
interaction with a dipole wave.
For these purposes the dipole axis, which is parallel to the rod axis,
has to be aligned along the PM's optical axis, as envisaged in
Fig.\,\ref{fig:setup}(b).
This should occur in a natural way due to the dominating longitudinal
electric field vector created by the focused radially polarized
trapping beam, which polarizes the rod material.
Recent calculations for elongated nanoparticles support this
expectation\,\cite{li2015}.

Moreover a characterization of the trapping dynamics at a
significantly reduced pressure should solidify the observed
increase of the magnitude of the trap frequencies in comparison to
microscope-objective based optical traps further.
Already at the current stage, the measured trap frequencies hint at
a trap stiffness unprecedented in free-space focusing.

Under vacuum conditions, the removal of heat from a trapped
particle is inefficient, since the collisions with air molecules are
less frequent.
Besides the well-known consequences for the particle motion, the
particle might melt due to heating by the absorption of the trapping
light.  
The latter effect could be especially severe in the tight focusing
condition found in a deep PM.
Recently, the melting of levitated particles by absorption of the
trap-laser light was observed in a low-pressure
environment\,\cite{millen2014}.
Therein, this effect has been proven to be especially significant for
micrometer-sized spheres and much less for nanospheres of 100\,nm diameter.
We therefore do not expect this effect for a single DR particle, whose
volume is about five to two orders of magnitude smaller, respectively.
Indeed, in combination with dedicated cooling
techniques\,\cite{gieseler2012,li2011},
reaching low temperatures 
for the trapped particle's motion\,\cite{rodenburg2016} appears to
be feasible with a dipole trap based on a deep parabolic mirror.

\section*{Funding Information}
G.L. acknowledges financial support from the European
Research Council via the Advanced Grant \lq PACART\rq .

\section*{Acknowledgments}
The authors acknowledge fruitful discussions with M.~Chekhova,
M.~Manceau and M.~Sytnyk.

%% \bibliographystyle{apsrev4-1}
%% \bibliography{tweezerBibliography}

\begin{thebibliography}{37}
\bibitem [{met()}]{methanolNote}%
  \BibitemOpen
  \href@noop {} {}\bibinfo {note} {Methanol was added to the DR solution which
  was originally toluene, because it is better adapted for spraying with the
  ultrasonic nebulizer. However, DRs are surrounded by alkyl chains that render
  them soluble in organic solvents, but polar solvents like methanol tend to
  promote nanoparticle aggregation.}\BibitemShut {Stop}%
\bibitem [{pow()}]{powerNote}%
  \BibitemOpen
  \href@noop {} {}\bibinfo {note} {From observing trapping powers as low as
  8.5\,mW it can be inferred that the trapped object has a polarizability that
  is roughly 5 times larger than calculated for a single CdS rod. Therefore, in
  first approximation one can deduce that the trapped object is a cluster
  consisting of $\ge$5 rods. However, it is questionable whether a cluster can
  still be described by a prolate spheriod. Modelling such a cluster as a CdS
  sphere, the volume of this sphere would amount to the volume of 7 rods with
  the size parameters given in Sec.\,\ref{sec:Concept}.}\BibitemShut {Stop}%
\end{thebibliography}

\end{document}